\documentstyle[aps,amsmath,prl,multicol,graphicx]{revtex}

\title{Optical excitations of a self assembled artificial ion}

\author{F. Findeis, M. Baier, A. Zrenner, M. Bichler, and G. Abstreiter}

\address{
Walter Schottky Institut, Technische Universit\"at M\"unchen\\ Am
Coulombwall, 85748 Garching, Germany}

\author{U. Hohenester and E. Molinari}

\address{
Istituto Nazionale per la Fisica della Materia (INFM) and
Dipartimento di Fisica\\ Universit\`a di Modena e Reggio Emilia,
Via Campi 213/A, 41100 Modena, Italy}

\date{January 16, 2001}
\begin{document}

\maketitle
%\vspace{1cm}

\begin{abstract}

By use of magneto-photoluminescence spectroscopy we demonstrate
bias controlled single-electron charging of a single quantum dot.
Neutral, single, and double charged excitons are identified in the
optical spectra. At high magnetic fields one Zeeman component of
the single charged exciton is found to be quenched, which is
attributed to the competing effects of tunneling and spin-flip
processes. Our experimental data are in good agreement with
theoretical model calculations for situations where the spatial
extent of the hole wave functions is smaller as compared to the
electron wave functions.

\end{abstract}


%\narrowtext
\begin{multicols}{2}

Semiconductor quantum dots (QDs) are often referred to as
artificial atoms. Different levels of neutral occupancies in QDs
have been obtained in the last years by power dependent optical
excitation. The associated few particle states have been
intensively studied by multi-exciton photoluminescence (PL)
spectroscopy and corresponding theoretical investigations
\cite{Zrenner:JCP,Deckel,Landin,Hawrylak:PRB,Hohenester:SSC,Findeis:SSC,Bayer:Nature}.
Occupancies with different numbers of electrons and holes result
in charged exciton complexes. In analogy to QDs with neutral
occupancy ---artificial atoms--- charged exciton complexes may be
considered as artificial ions. In the field of low dimensional
semiconductors charged excitons have been observed first in
quantum well structures \cite{Kheng}. In QDs, charged excitons
have been studied in inhomogeneously broadened ensembles by PL
\cite{Schmidt} as well as in interband transmission experiments
\cite{Warburton:PRL}, and recently also in single, optically
tunable QDs \cite{Hartmann} as well as electrically tunable
quantum rings by PL \cite{Warburton:Nature}.

Few-particle theory predicts binding energies for charged QD
excitons in the order of some meV \cite{Hartmann,Wojs}. This
allows for the controlled manipulation of energetically well
separated few particle states under the action of an external gate
electrode. Discrete and stable numbers of extra charges are
thereby possible via the Coulomb blockade mechanism. In future
experiments and possible applications the resonant optical
absorption and emission of such systems is expected to be tunable
between discrete and characteristic energies. Moreover such
few-particle systems are expected to exhibit an interesting
variety of spin configurations, which can be controlled by
external magnetic field, gate induced occupancy, and spin
selective optical excitation.

In the present contribution we present for the first time
experimental and theoretical results on the gate-controlled
charging of a single InGaAs QD with an increasing number of
electrons probed by magneto-PL.

For controlled charging of individual QDs a special electric field
tunable n-i structure has been grown by molecular beam epitaxy.
The In$_{0.5}$Ga$_{0.5}$As QDs are embedded in an i-GaAs region 40
nm above an n-doped GaAs layer ($5\times10^{18}cm^{-3}$) which
acts as back contact. The growth of the QDs is followed by 270 nm
i-GaAs, a 40 nm thick Al$_{0.3}$Ga$_{0.7}$As blocking layer, and a
10 nm i-GaAs cap layer. As a Schottky gate we use a 5 nm thick
semitransparent Ti layer. The samples were processed as
photodiodes combined with electron beam structured shadow masks
with apertures ranging from 200 nm to 800 nm. Schematic overviews
of the sample and the band diagram are shown in Figs. 1 a and b.
The occupation of the QD with electrons can be controlled by an
external bias voltage $V_B$ on the Schottky gate with respect to
the back contact. For increasing $V_B$ the band flattens and the
electron levels of the QD are shifted below the Fermi energy of
the n-GaAs back contact, which results in successive single
electron charging of the QD. In our experiments excitons are
generated optically at low rate and form charged excitons together
with the $V_B$ induced extra electrons in the QD. We used a HeNe
laser (632.8 nm) for optical excitation and a cooled CCD camera
for detection of the PL. The sample was mounted in a confocal
low-temperature, high magnetic field microscope
\cite{Zrenner:Physica}.

In Fig. 2 a and b we present PL spectra as a function of $V_B$ in
the range of -550 mV to +400 mV corresponding to electric fields
of 37.5 kV/cm to 11.1 kV/cm. The PL intensity is displayed as gray
scale plot for B=0 T in Fig. 2 a and for B=12 T in Fig. 2 b. As a
function of $V_B$ we find a series of lines with discrete jumps in
the emission energy. Those lines are assigned to radiative s-shell
transitions of neutral ($X^0$), single charged ($X^-$), and double
charged ($X^{2-}$) excitons, as marked in Figs. 2 a and b and
discussed in the following. For B=0 T (Fig. 2 a) and large
negative $V_B$ ($V_B
<-0.5~V$) the electron levels in the QD are far above the Fermi
energy of the n-GaAs back contact and the QD is electrically
neutral. Optically generated excitons can relax into the QD, but
before radiative recombination ($\tau \sim 1 ns$) the carriers
tunnel out of the QD due to the high applied electric field. For
$V_B\simeq -0.5~V$ the QD is still uncharged, but in the smaller
electric field radiative recombination gets more likely and the
$X^0$ emission line appears at 1307 meV. The weak satellite about
0.4 meV below is tentatively explained in terms of QD asymmetry
\cite{Gammon:PRL}. With increasing $V_B$, the $X^0$ line shifts to
higher energies due to the quantum confined Stark effect in the
decreasing electric field. For $V_B =-0.35~V$ a new emission line
appears below the $X^0$ line at 1302.5 meV, indicating the static
occupation of the QD with one electron. The $X^-$ binding energy
with respect to the $X^0$ is determined to $\Delta E_{X^-}= 4.6~
meV$ by the measured difference in emission energies. For
$-0.35V<V_B<0V$ the $X^0$ and $X^-$ line coexist, which is a
consequence of the statistical nature of non-resonant optical
excitation. In the presence of one extra electron, the capture and
subsequent decay of an electron hole pair leads to the emission of
a $X^-$ photon. If only a single hole is captured we expect a
$X^0$ photon, and if a single electron is captured we expect no
photon, but instead electron back transfer to the $n^+$ region. At
$V_B=0~V$ the QD is charged with a second electron. This leads to
the appearance of two new characteristic emission lines (marked in
Fig. 2 a and b), which are assigned to the $X^{2-}$ decay. The
main line of the $X^{2-}$ emission appears only 0.3 meV below the
$X^{-}$ line, whereas a much weaker satellite peak appears 4.6 meV
below the main line at 1298.1 meV. The appearance of two emission
lines is characteristic for the $X^{2-}$ decay. The energy
difference between the two $X^{2-}$ lines corresponds to the
difference in the s-p exchange energies between the two possible
final states with parallel or antiparallel spin orientation of the
two remaining electrons \cite{Hawrylak:PRB}. Again $X^{2-}$ and
$X^-$ can be observed simultaneously over a certain range in
$V_B$. At $V_B>0.19 V$ only one broad emission line remains. This
indicates filling of the wetting layer (WL) states with electrons.
Here, weakly confined electrons are interacting with the carriers
in the QD, causing a broadening of the detected s-shell decay in
the QD. A rough estimation of the $V_B$ increment needed to bring
the WL states below the Fermi energy (starting from the onset of
the $X^-$ line), taking charging energy and the electrostatics of
the structure into account \cite{Drexler:PRL}, results in a
reasonable value of $\Delta V_B =530~mV$ in good agreement with
our experimental data.

% Theory ----------------------------------------------------

For a quantitative analysis of the experimental results we
performed theoretical model calcualtions. Following the approach
presented in Ref.~\cite{navarez:00}, we assume for electrons and
holes, respectively, a confinement potential which is parabolic in
the $(x,y)$-plane and box-like along $z$; despite its simplicity,
such confinement is known to mimic the most important
characteristics of In$_x$Ga$_{1-x}$As dots, and has recently
proven successful in comparison with experiment
\cite{Bayer:Nature,Hawrylak:PRL}. We take 5 nm for the well width
in $z$-direction and $\hbar\omega_0^e=30$ meV
($\hbar\omega_0^h=15$ meV) for the electron (hole) confinement
energies due to the in-plane parabolic potential; material
parameters are computed according to Refs.~\cite{stier:99,valence}
(note that with these values we also well reproduce the $\sim$40
meV splitting between the $1s$ and $1p$ shells measured at higher
photo-excitation powers). Finally, because of the strong quantum
confinement in $z$-direction we safely neglect minor effects due
to the applied external electric field. PL spectra for $X^0$,
$X^-$, and $X^{2-}$ are computed within a direct-diagonalization
approach accounting for all possible e-e and e-h Coulomb
interactions \cite{Hartmann,details}

The black lines in Fig.~3 show the dependence of the luminescence
spectra of the $X^0$, $X^-$, and $X^{2-}$ lines as function of the
extension $L_0^h$ of the hole wavefunction (for definition see
figure caption). In accordance with experiment and related work
\cite{Hartmann,Warburton:Nature} we find (for $L_0^h\leq L_0^e$)
in the PL spectra: a red-shift of the charged-exciton emission
peak $X^-$; a further, much smaller, red-shift of the main
$X^{2-}$ emission peak, which is accompanied by a weak satellite
peak at even lower photon energy (see arrows). A quantitative
comparison between theory and experiment, however, reveals that
for the same extension of electron and hole wavefunctions
($L_0^e=L_0^h$) the calculated charged-exciton binding of 1.5 meV
is by a factor of $\sim$3 smaller than the measured value; we
checked that this finding does not depend decisevly on small
modifications of the chosen dot and material parameters. As can be
inferred from Fig.~3, the $V_B$ binding energy is a quite
sensitive measure of the relative extension of electron and hole
wavefunctions: Decreasing $L_0^h$ and keeping all other parameters
($\omega_0^e$, $\omega_0^h$, $L_0^e$) fixed, one clearly observes
in Fig.~3 an increased binding (moderate parameter changes turn
out to have no impact on this general trend; see, e.g., gray
lines).

We thus conclude that the large experimental $X^-$ binding energy
of 4.6 meV can only be explained by more localized hole
wavefunctions, which we attribute to effects of heavier hole
masses and of possible piezoelectric fields. Additional evidence
for this interpretation comes from the ratio between the exchange
splitting of the two $X^{2-}$ lines and the $X^-$ binding energy.
The experimentally observed exchange splitting of about 4.6 meV
between the weak low energy and intense high energy branch of the
$X^{2-}$ line is found to be approximately equal to the $X^-$
binding energy. From our theoretical investigations such a
scenario can only be obtained if the hole confinement is
considerably stronger than the electron confinement.

Finally we discuss the magnetic field dependence shown in Fig. 2 b
for B=12 T. From comparison with the B=0 T data (Fig. 2 a) it is
clear that single electron charging versus $V_B$ is mostly
unaffected by magnetic field. The centers of s-shell emission for
$X^0$, $X^-$, and $X^{2-}$ are shifted to higher energies due to
diamagnetic shift and each emission line splits into two lines,
separated by the Zeeman energy. The observed weak differences
between the Zeeman energies of the $X^0$ and $X^-$ lines seem to
indicate, that the spin orientation of an extra electron in the QD
does not change during the radiative decay from the s-shell.

In the following we concentrate on the asymmetry in the
PL-intensities of the two Zeeman branches of the $X^-$ line for
$-0.35~V<V_B<-0.13~V$ (see Fig. 2 b). This asymmetry is observed
only for the $X^-$ line, not for the $X^0$ and $X^{2-}$ lines. The
explanation of this phenomenon involves spin polarization, Pauli
blocking, and state selective tunneling as summarized in Fig. 4 a
and b. At B=12 T a single electron in a QD is spin polarized in
thermal equilibrium. The optical excitation of excitons can happen
with two different spin orientations, which results in the states
shown in Fig. 4 a and b. Due to Pauli blocking in the conduction
band, parallel electron spin orientation leads to a metastable
triplet state with one electron in the s-shell and one in the
p-shell (see Fig. 4 a). If the tunneling time from the p-shell to
the continuum is shorter than the electron spin-flip time, an
electron is lost and we end up with a neutral exciton and hence
with a contribution to one Zeeman component of the $X^0$ line,
i.e. we lose one Zeeman component of the $X^-$ decay. The
introduction of an exciton with opposite spin orientation,
however, produces a singlet state as shown in Fig. 4 b. The
radiative decay of this configuration contributes to the other
Zeeman component of the $X^-$ line. Our findings also imply that
the associated heavy hole spin flip time should be at least of the
same order or longer than the e-h lifetime. Long spin-flip times
and conservation of the exciton spin within the exciton lifetime
has been reported already earlier for zero-dimensional systems at
high magnetic fields \cite{Heller:PRB}. The suppression of one
$X^-$ Zeeman component persists over a $V_B$ range of 0.22 V,
which translates to an energy shift of the QD states with respect
to the n-GaAs Fermi energy of 28 meV (about to conduction band s-p
separation). For $V_B>-0.13~V$ both Zeeman components of the $X^-$
line are recovered as a consequence of the increased tunneling
time, which allows now for the competing spin flip and relaxation
process to the s-shell. For the $X^0$ and $X^{2-}$ lines a
quenching of Zeeman lines is not expected and also not observed,
since the p-shell is either never ($X^0$) or always ($X^{2-}$)
populated, regardless of the spin orientation of the optically
excited e-h pair.

In summary, we have demonstrated bias controlled charging of a
single QD in magneto-optic PL experiments. The few-particle
interaction energies determined experimentally for the $X^-$ and
$X^{2-}$ states are found to be in good agreement with our
theoretical model for situations where the spatial extent of the
hole wave functions is smaller as compared to the electron wave
functions. Spin polarization and lifting of the Zeeman degeneracy
at high magnetic fields allows further access to so far unexplored
spin dependent properties of few particle states. The suppression
of one Zeeman component of the $X^-$ decay is explained in terms
of state selective tunneling from a spin triplet configuration.

This work has been supported financially by the DFG via SFB 348,
by the BMBF via 01BM917, in part by INFM through PRA-99-SSQI, and
by the EU under the TMR Network ``Ultrafast Quantum
Optoelectronics'' and the IST programme ``SQID''.

\begin{figure}
%\centerline{\includegraphics[width=0.75\columnwidth]{fig1.png}}
%\vspace*{0.5cm}
\caption{(a) Photodiode combined with a near field shadow mask.
(b) Schematic band diagram of the structure for zero bias.}
\label{fig1}
\end{figure}

\begin{figure}
%\centerline{\includegraphics[width=0.75\columnwidth]{fig2.png}}
%\vspace*{0.5cm}
\caption{Gray scale plot of PL intensity as a function of PL
energy and VB. The series of lines can be assigned to emissions
from s-shell transitions of $X^0$, $X^-$, and $X^{2-}$. At zero
magnetic field (a) and at B=12 T (b)} \label{fig2}
\end{figure}

\begin{figure}
%\centerline{\includegraphics[width=0.85\columnwidth]{fig3.pdf}}
\caption{ Dependence of luminescence spectra on the extension of
the hole wavefunction. $L_0^i=1/\sqrt(m_i\omega_0^i)$ ($i=e,h$) is
a characteristic length scale of the parabolic confinement, with
$m_e$ ($m_h$) the electron (hole) mass and $\hbar\omega_0^e$
($\hbar\omega_0^h$) the confinement energies due to the in-plane
parabolic potential. Black lines:  $\hbar\omega_0^e=30$ meV
($L_0^e=7.5$ nm) and $\hbar\omega_0^h=15$ meV; the thickness of
each line corresponds to the oscillator strength of the
corresponding transition (photon energy zero given by the $X^0$
energy for same extension of electron and hole wavefunctions,
i.e., $L_0^h=L_0^e$). Gray lines: $\hbar\omega_0^e=35$ meV
($L_0^e=7$ nm) and $\hbar\omega_0^h=10$ meV. } \label{fig3}
\end{figure}

\begin{figure}
%\centerline{\includegraphics[width=0.85\columnwidth]{fig4.png}}
%\vspace*{0.5cm}
\caption{Triplet (a) and singlet (b) electron configurations for
the $X^-$. The triplet state is subjected to tunnel decay from the
p-shell.} \label{fig4}
\end{figure}

\end{multicols}


\begin{thebibliography}{99}
\bibitem{valence} We use $m_z^h=1/(\gamma_1-2\gamma_2)$ and
$m_\|^h=1/(\gamma_1+\gamma_2)$ for the hole mass in the $z$ and
lateral directions, with $\gamma_i$ the Luttinger parameters.
\end{thebibliography}
\end{document}